%% file: chi.tex
\documentclass[sigconf, authorversion]{acmart}
\usepackage{longtable}
\usepackage{graphicx}
\usepackage{soul}
\pdfoutput=1
\AtBeginDocument{%
  }

\copyrightyear{2024}
\acmYear{2024}
\setcopyright{acmlicensed}\acmConference[CHI '24]{Proceedings of the CHI Conference on Human Factors in Computing Systems}{May 11--16, 2024}{Honolulu, HI, USA}
\acmBooktitle{Proceedings of the CHI Conference on Human Factors in Computing Systems (CHI '24), May 11--16, 2024, Honolulu, HI, USA}
\acmDOI{10.1145/3613904.3642645}
\acmISBN{979-8-4007-0330-0/24/05}

\newcommand*{\marked}[1]{#1} 

\begin{document}

\title[Supportive Fintech for Bipolar Disorder]{Supportive Fintech for Individuals with Bipolar Disorder: Financial Data Sharing Preferences to Support Longitudinal Care Management}

\author{Jeff Brozena}
\authornote{Both authors contributed equally to this research.}
\email{brozena@psu.edu}
\orcid{0000-0001-7621-1566}
\affiliation{%
  \institution{Penn State University}
  \city{University Park}
  \state{PA}
  \country{USA}
}

\author{Johnna Blair}
\authornotemark[1]
\email{jlb883@psu.edu}
\orcid{0000-0001-9579-9393}
\affiliation{%
  \institution{Penn State University}
  \city{University Park}
  \state{PA}
  \country{USA}
}

\author{Thomas Richardson}
\orcid{0000-0002-5357-4281}
\email{t.h.richardson@soton.ac.uk}
\affiliation{%
  \institution{University of Southampton}
  \country{United Kingdom}}

\author{Mark Matthews}
\orcid{0000-0001-8862-5141}
\email{mark.matthews@ucd.ie}
\affiliation{%
  \institution{University College Dublin}
  \city{Dublin}
  \country{Ireland}
}

\author{Dahlia Mukherjee}
\orcid{0000-0003-1007-2094}
\email{dmukherjee@pennstatehealth.psu.edu}
\affiliation{%
 \institution{Penn State College of Medicine}
 \city{Hershey}
 \state{PA}
 \country{USA}}

\author{Erika F. H. Saunders}
\orcid{0000-0001-7222-0828}
\email{esaunders@pennstatehealth.psu.edu}
\affiliation{%
  \institution{Penn State College of Medicine} 
  \city{Hershey}
  \state{PA}
  \country{USA}}

\author{Saeed Abdullah}
\orcid{0000-0002-4371-8173}
\email{saeed@psu.edu}
\affiliation{%
 \institution{Penn State University}
 \city{University Park}
 \state{PA}
 \country{USA}}

\renewcommand{\shortauthors}{Brozena et al.}

\begin{abstract}
Financial stability is a key challenge for individuals living with bipolar disorder (BD). Symptomatic periods in BD are associated with poor financial decision-making, contributing to a negative cycle of worsening symptoms and an increased risk of bankruptcy. There has been an increased focus on designing supportive financial technologies (fintech) to address varying and intermittent needs across different stages of BD. However, little is known about this population’s expectations and privacy preferences related to financial data sharing for longitudinal care management. To address this knowledge gap, we have deployed a factorial vignette survey using the Contextual Integrity framework. Our data from individuals with BD (N=480) shows that they are open to sharing financial data for long term care management. We have also identified significant differences in sharing preferences across age, gender, and diagnostic subtype. We discuss the implications of these findings in designing equitable fintech to support this marginalized community.
 

\end{abstract}



\maketitle

\section{Introduction}
Bipolar disorder (BD) is a serious mental illness characterized by periods of mania and depression. A chronic illness with no known cure, BD requires life-long management of symptoms varying in intensity through time. BD ranks as the sixth leading cause of disability globally \citep{murrayglobal1996} and carries significant macroeconomic impacts. A 2015 analysis estimated its total costs in the US economy to be \$202.1 billion, amounting to \$81,559 per individual \citep{cloutiereconomic2018, bessonovaEconomic2020}. While BD has been associated with an increased capacity for creativity \citep{greenwoodCreativity2020}, its disruptive impacts are often significant and long-lasting.

Financial difficulties are so frequently associated with symptomatic periods of BD that the American Psychiatric Association includes anomalous spending or otherwise risky financial behaviors among its diagnostic criteria \citep{americanpsychiatricassociationDiagnostic2013}. A recent population-scale study found that individuals with BD type I were 50\% more likely to declare bankruptcy than the healthy comparison group \citep{nauAssessment2023}. Money-related factors have been shown to contribute to a worsening of mental health symptoms, including lower self-esteem and debt-related anxiety \citep{richardsonFinancial2021}. 

We believe these challenges pose an unmet call to design data-driven tools to support financial stability throughout the phases of BD. Prior work has demonstrated that financial technologies can be appropriately situated within the lives of individuals with mental illness \citep{barrospenaFinancial2021, blairFinancial2022}, including in ways that support third-party collaboration \citep{barrospenaPick2021}. Open banking technologies now afford access to financial data at the granularity necessary to create systems that can deliver timely, personalized support for these highly individualized needs. However, we do not yet know if individuals with BD are comfortable sharing their financial data to support longitudinal care management. 

To this end, we deployed a large-scale international survey (N=480) to explore financial data sharing preferences of individuals with BD. We used a factorial vignette design to understand sharing preferences across different contexts: how much financial information is shared, with whom it is shared, and for what purposes. To complement these insights, we also asked participants about their current financial management strategies and how their care partners have helped them manage spending behaviors. In this work, we present both quantitative and qualitative findings to highlight the complex needs and attitudes toward financial data sharing that exist within this population. 

Overall, participants were most comfortable using financial data for self-management (i.e., sharing data with only themselves), followed by sharing with clinicians, and then family. However, there were some demographic differences in these sharing preferences. In particular, those who were married or had children were more likely to be comfortable sharing financial data with families. Conversely, women were considerably less likely to share financial data with family. The survey data also showed that the majority of participants involved their care partners in managing their finances, to varying degrees, from having full control of their bank accounts to offering occasional financial advice.

Based on these insights, we establish that individuals with BD are interested in using financial data for long-term BD management. However, there are complex privacy needs and risks that must be considered when designing support systems built on financial data sources. We also discuss design recommendations for using financial data to support long-term BD management. More specifically, financial data sources show promise to support BD self-management, improve financial decision-making and literacy, assist with care partner collaboration, and integrate into existing clinical practices. 

\section{Related Work}

\subsection{Bipolar Disorder and Financial Instability}

\marked{BD is a chronic condition characterized by reoccurring episodes of depression and mania or hypomania \cite{americanpsychiatricassociationDiagnostic2013}. Depressive episodes, much like major depression, involve low melancholic mood states \cite{ketter2010diagnostic}. Conversely, mania is a period of elevated mood that features heightened activity and impulsivity. While hypomania shares many characteristics with mania, it is associated with less severe impairment \cite{guzman2021clinical}. BD can be characterized into major subtypes including BD I, BD II, and cyclothymia. BD I and BD II are differentiated by the occurrence of manic episodes. BD I involves at least one episode of mania along with depressive episodes \cite{guzman2021clinical, phillips2013bipolar}. BD II, on the other hand, involves depressive and hypomanic episodes, but without any episode of mania. Cyclothymia involves periods of mood changes, but are less extreme \cite{phillips2013bipolar, parker2012cyclothymia}. For all subtypes, the length and frequency of mood episodes can vary considerably across individuals.}

\marked{Financial instability is a critical concern in BD. Indeed, the American Psychiatric Association considers risky or impulsive financial behavior to be a diagnostic criterion of BD \cite{americanpsychiatricassociationDiagnostic2013}. Impulsivity is common during manic and hypomanic episodes \cite{fletcherHighrisk2013}, which can lead to poor financial decision-making. On the other hand, depressive episodes have been associated with cognitive impairments resulting in reduced attention, working memory, and executive functioning \cite{marazzitiCognitive2010}. These issues can lead to challenging financial consequences during depressive periods \cite{richardsonrelationship2017}. BD is also associated with increased risk-taking behaviors \cite{ramirez-martinImpulsivity2020, fletcherHighrisk2013}, which can lead to financial instability. Indeed, the risk of bankruptcy is 1.5 times higher in BD compared with the general population \cite{nauAssessment2023}.}

\marked{The resultant financial instability can cause significant challenges for the care partners \citep{reinaresPsychosocial2014} and family members \citep{targumFamily1981}. Individuals with BD and their care partners often need to establish collaborative strategies to minimize the impact of problematic financial decision-making over different stages of BD \citep{richardsonFinancial2021}. However, this requires continuous assessment of financial behaviors, which can be a challenging task for both individuals with BD and their care partners. Prior work has relied on self-reported data to assess the relationship between mental health states and financial decision-making \cite{richardsonFinancial2021}. However, these approaches are not suitable for longitudinal monitoring nor timely interventions required to maintain financial stability in BD \cite{blairFinancial2022}}.

\subsection{Open Banking and Supportive Financial Technology Design}

\marked{In recent years there has been significant progress in financial data access. Patterns of social interaction surrounding financial transactions have been explored in "moneywork" literature \citep{perryMoneywork2018}. "Moneywork" can be defined as the tasks and practices individuals engage with that surround money and making payments\cite{perryMoneywork2018}. The moneywork framing has been extended to the study of supportive behaviors for individuals with disabilities \citep{kameswaranCash2019}. These behaviors included ``articulation work'', or supportive, pre-transactional activities, which made cash available to individuals with vision impairments to conduct their business endeavors. Participants' perceived necessity for aid was found to increase trust between these individuals and their collaborators. Moneywork has also been applied and extended to supportive third-party access in mental health contexts \citep{barrospenaPick2021}. Researchers implemented the UK Open Banking API to create and deploy a third-party financial access tool, recruiting 14 individuals who self-identified as living with a mental health condition. Each individual chose a ``trusted ally'' who received notifications when specific types of transactions took place as a means to support financial collaboration. Marginalized groups may experience additional barriers to banking access or distrust towards banking institutions, leading users to develop their own strategies for adopting fintech\cite{cunningham2022cost}. Additionally, existing fintech can be difficult to navigate or even help hold power imbalances in the case of financial abuse \cite{belliniPaying2023}. This suggests a need for new fintech systems that are empathetic to the experiences of more marginalized users.}

\marked{Banking institutions have begun to offer supportive features for individuals with mental illness who require support \emph{some} of the time \citep{farrWhy2019}. Third-party supportive behaviors have been conceptualized as \emph{monitoring behaviors}, such as view-only access to financial statements, and \emph{controlling behaviors}, such as payee relationships or self-imposed spending limits. Prior work on financial intervention design has shown that goal-oriented interventions are among the most effective and engaging \citep{jimenez-solomonPeersupported2016} and that the inclusion of self-prescribed goals or parameters in algorithmic interventions are among the most strongly called for \citep{farrWhy2019}. Institutions might consider adding frictions to spending such as self-imposed limits to debit card usage, or the ability to set a ``cooling off'' period for transactions of a certain dollar amount or made during specified hours. The concept of frictional microboundaries exists elsewhere in HCI literature as a means of encouraging ``mindful reflection'' \citep{coxDesign2016} in order to slow down the actions of a user as a prompt for individuals to engage in more intentional actions. However, there is a lack of established privacy norms that are needed to effectively design supportive fintech for this specific population.}

\subsection{Privacy Norms for Health Data Sharing}

\marked{Personal health informatics have provided new opportunities for users to engage with their personal data, reflect on behavioral patterns, and decide to act on selected behaviors to meet their health goals \cite{epstein2018everyday}. This data can provide new resources for self-management or as a collaborative tool. It also allows users to share information with others for social support or accountability towards improved health habits \cite{lepore2021supportive} or the management of chronic conditions \cite{tahsin2022examining}. With the rising use of personal health informatics, there has also been a significant effort to understand the needs and concerns related to sharing personal health data. For instance, prior work has shown that users may desire to exclude certain information from being collected, as exemplified by momentarily disabling their data capture over time \citep{gouveiaActivity2018}. Similarly, sharing preferences can vary considerably depending on the data recipient \citep{epsteinFinegrained2013a}. Prior work has also found that data sharing can lead to unintended consequences, including negative impact on social relationships \citep{epsteinExamining2017}. For example, inconsistencies or anomalies in shared health data run the risk of misinterpretation by others, signaling the need for annotation, justification, or explanation in possible designs \citep{leeReflecting2011, vandenbergheSelf2018}.}

\marked{The privacy needs of marginalized populations might differ considerably given the increased risk of potential harms \cite{sannon2022privacy}. For instance, despite the opportunity for improved communication and care that can result from sharing information, individuals with HIV can be reluctant to share data and engage in information disclosure \cite{claisse2023perspectives}. Similarly, individuals with BD might also have complex privacy needs and concerns regarding data sharing. Morton et al. \cite{morton2021use} found that while individuals want additional tools to support BD management, data security and privacy remains a top concern for them. Petelka et al. \cite{petelkaBeing2020} also reported how privacy concerns impact intentional sharing of BD-related experiences with others. They identified different factors that can determine how and when individuals decide to reveal or conceal their BD status to others, such as a need for transparency within care networks, potential risks for exposure, or a perceived familiarity with BD \cite{petelkaBeing2020}.}

\marked{Previous research has explored privacy standards for personal informatics in BD management, but little has been done to understand privacy requirements in financial data sharing. We consider this to be a serious knowledge gap given the critical need for maintaining financial stability and collaborative management for this population. Moreover, financial instability following poor decision-making can be highly stigmatized and it can lead to persistent feelings of regret, guilt, and shame in individuals with BD \cite{richardsonFinancial2021}. As such, they might have complex and varied needs and concerns toward sharing financial information. Identifying the privacy norms and expectations in this context is essential to develop supportive fintech for individuals with BD. Our work aims to address this important knowledge gap by focusing on the following research questions:}

\begin{enumerate}

    \item How do individuals with BD feel about using their financial data for self-management of BD?
    \item How do individuals with BD feel about sharing financial data with others?
    \item How do their data sharing attitudes change regarding a) recipient, b) contexts of use, and c) data types? 
    \item How do individuals with BD currently involve care partners in their financial management strategies?
\end{enumerate}

\section{Methods}

This study focuses on identifying privacy norms, preferences, and expectations related to financial data sharing for longitudinal care management in BD. \marked{Given these goals, we decided to use an online survey so that we can collect data from a large sample across different geographical regions. The use of an online survey specifically allowed us to explore how attitudes, experiences, and privacy norms might vary across different subgroups}. In the following sections, we describe the survey design, data collection process, and analysis steps. 

\subsection{Survey Design}
We used the Contextual Integrity (CI) framework \citep{nissenbaumPrivacy2004} for the survey. Recent studies have successfully used the CI framework to assess privacy norms and concerns regarding health data sharing \citep{silberVignette2021,utzApps2021}. The CI framework theorizes that privacy norms surrounding information transfers are determined by both i) appropriateness of information sharing in a given context; and ii) recipients of shared information. Following the CI framework, the survey used a factorial vignette design \citep{finchVignette1987}. We asked survey participants to provide privacy ratings of hypothetical scenarios. Each scenario can include multiple factors representing individual elements of the CI framework (e.g., data types, context of uses, and data recipients). The factorial vignette approach systematically varies these factors to assess which contextual factors might impact privacy norms and data sharing preferences. Furthermore, it allows determining relative importance of these factors across different subgroups.

\hypertarget{section-vignettes}{%
\subsubsection{Vignette Factors}\label{section-vignettes}}

\marked{To design the vignettes, we used prior work to identify potentially relevant factors including type of financial account, granularity of collected financial data, duration of data storage, and primary and secondary contexts of use \cite{reinaresrole2016, blairFinancial2022, richardsonFinancial2021}. The authors then iteratively selected contextual factors that are highly relevant for this population, while balancing the participant burden and survey length.}  These selected factors included:  (1) actors (recipients of information); (2) contexts of data use; and (3) data granularity. 

\marked{We also identified relevant levels for these factors. We included three different actors --- clinician, family, and themselves. We selected these stakeholders as actors given their common involvement in long-term BD care \cite{reinaresrole2016, richardsonFinancial2021} as well as financial \cite{barrospenaPick2021} and healthcare \cite{murnanePersonal2018} management. We also explored how different levels of granularity might impact financial data sharing attitudes. To identify relevant granularity levels, we used findings from prior work \cite{barrospenaFinancial2021, blairFinancial2022, lewisFollow2019} as well as existing capabilities of fintech platforms \cite{plaid}. This resulted in three levels reflecting different financial data granularity: i) timing and amount, ii) timing and category (e.g., travel, utilities), and iii) all purchase details in a given transaction. Furthermore, we identified relevant use contexts following prior work on financial behaviors and needs of this population \cite{blairFinancial2022, richardsonFinancial2021}.}

The resultant survey used 3 factors each containing 3 levels (total of 27 vignettes). Table~\ref{tbl-factors} lists these factors and their respective levels. Each vignette was created using the following template: ``An app accesses {[}Granularity{]} so it can {[}Context{]}. This app will share these insights with {[}Recipient{]}.'' We chose a full-factorial design for the survey with each participant rating all vignettes. For each vignette, participants rated their level of comfort when sharing financial data on a scale of 0--10, where 0 is ``extremely uncomfortable'' and 10 is ``very comfortable.''


\input{tables/factors}

\subsection{Survey Questionnaire}

The following section details the questions used in the survey. This survey questionnaire can be found in our supplementary material. 


\subsubsection{Financial Environments}

We asked about their employment status and information related to their annual individual income excluding welfare or benefits, whether they had access to a financial institution, how often they check their account balances, and what methods they use to review their spending behaviors. We also asked individuals to self-report their debt in their local currency. However, due to high levels of missing data, we did not include this section in our analysis. Additionally, participants rated their agreement towards a series of statements regarding financial hardship, worry, and anxiety. We also collected data about the perceived frequency, volume, and category of their online transactions when using a credit or debit card, amount and nature of purchases they made with cash, and their usage patterns of money-related digital technologies.

\subsubsection{BD-related Spending Behaviors}

Following this, we prompted users to consider whether their spending changes during manic or depressive episodes, in what specific ways and about their specific goals for these types of spending. We also asked participants whether they have ever (1) considered bankruptcy or (2) had ever declared bankruptcy due to large purchases or impulsive spending occurring during a manic episode.

\subsubsection{Financial Management Strategies}

To better understand their current practices of self-management and collaborative management of impulsive financial behaviors, we asked individuals whether they actively attempt to reduce or prevent impulsive spending, what strategies they make use of towards those goals, if and how they have involved family or friends to help prevent impulsive spending, and whether they would be willing to rely on technology or their creditors to help impede overspending.

\subsubsection{Vignettes}

As described in Section \ref{section-vignettes}, we then concluded the survey by presenting participants with a total of 27 hypothetical scenarios to determine their level of financial data sharing preferences across three dimensions of the CI framework---actor (recipient), context of use, and granularity (data type).


\subsubsection{Demographics}
\marked{The survey collected demographic information from participants including age, gender, race and ethnicity, education, and marital status. The data allowed us to explore personal characteristics associated with different sharing attitudes.}


\subsection{Survey Procedures}
\subsubsection{Pre-testing}
\marked{Prior to deployment, we tested our survey in multiple steps to ensure effective data collection. We first conducted internal testing focusing on readability, ordering effects, and clarity. The survey was then reviewed by clinicians in our team. We then collected data from a small pilot (N=12) to identify any potential concerns.}

\subsubsection{Deployment}

We deployed the survey in the United States, Ireland, and the United Kingdom between July 2022 to May 2023. We focused on these countries due to institutional access available to authors. Furthermore, we wanted to explore potential differences in privacy norms and expectations across different geographical regions. We shared the survey through social media, \marked{including Twitter, LinkedIn, and BD-related Facebook groups, and the distribution channels of several international organizations including the Depression and Bipolar Support Alliance (DBSA), Bipolar UK, and CrestBD, as well as local clinics}. 

\marked{The selection criteria included self-reported bipolar diagnosis, which is consistent with prior work on this population \citep{matthewsdoubleedged2017, hindleyWhy2019}. Following informed consent steps, the survey asked respondents whether they had received any BD-specific diagnosis from a clinician. Individuals without a BD diagnosis were excluded from the survey. At the end of the data collection, 10 participants were randomly selected for a gift card equivalent to \$50.}

\subsubsection{Ethics}
The study was approved by the Institutional Review Board (IRB) and Ethics committees at the relevant organizations. Furthermore, we followed relevant data protection guidelines including the General Data Protection Regulation (GDPR) from the European Union. \marked{Prior to any survey questions, potential respondents were provided an informed consent page. This detailed the broad goals of the survey and the types of questions to expect throughout (e.g., financial behaviors and debt in relation to BD episodes), the potential risks of participating, and study contact information. Individuals were told that their participation was voluntary and that they may refrain from responding or end the survey at any time should they feel uncomfortable.  At the end of the survey, participants were provided a briefing statement that included further information about the study aims and a list of country-specific resources for further psychological support if the participants became distressed by the nature of this survey.} \marked{To reduce privacy risks, the survey minimized collection of identifiable information from participants. The data storage and analysis steps followed the protocol approved by the relevant IRB and Ethics committees.}


\hypertarget{data-analysis}{%
\subsection{Data Analysis}\label{data-analysis}}

To better understand the complexity of financial data sharing preferences, we took a mixed methods approach to analysis. We conducted statistical analyses on participant vignette responses to determine attitudes toward the hypothetical financial data sharing scenarios. We used a qualitative, thematic analysis to understand the current ways participants actively engaged with their care partners to manage their financial decisions.

\subsubsection{Data Preparation}
\marked{Due to the sensitive topics discussed in this survey, we allowed respondents to skip questions. 67\% participants (N=324) answered all questions in the survey. The remaining 158 respondents skipped at least one question. Questions related to debt amounts and type of debt were the most commonly skipped questions. Our analysis included survey responses from all participants that met the inclusion criteria.}

\subsubsection{Statistical Analysis}

We used R for the data analysis. The vignette survey data has a hierarchical structure given each respondent answered all vignettes. We used multilevel modeling (\texttt{lme4}) to account for the nested data structure. The resulting dataset was preprocessed to maintain a ``long'' format compatible with the requirements of \texttt{lme4}. Our response variable (\texttt{y} in the formula given in supplementary materials) is the numerical rating for each vignette. \marked{We used maximum likelihood estimation \citep{endersPerformance2001, endersMissing2023} to account for missing data.}


\hypertarget{model-selection}{%
\paragraph{Model Selection}\label{model-selection}}

We followed best practice guidance \citep{zuurMixed2009, meteyardBest2020, baguleyStatistical2022} during model creation, selection, and validation. Specifically, we modeled unconditional means using maximum likelihood (ML) in order to assess the variance of random effects, initially chosen based on sampling unit (e.g., participant and vignette item). We defined a maximal model to include closed-ended factors for analysis and set random intercepts per respondent. We used \texttt{buildmer} \citep{voetenUsing2019} to perform backwards variable selection. Model selection was performed by iteratively removing non-significant factors based on the Bayes information criterion (BIC) of the resulting models. BIC has been shown to choose models that are more parsimonious than alternative methods \citep{neathBayesian2012} during model selection, making it an appropriate choice for our exploratory analysis. Visual inspection of residual plots did not show obvious deviations from homoscedasticity or normality. The final maximal model formula can be found in our supplementary materials. 

\hypertarget{estimated-marginal-means}{%
\paragraph{Estimated Marginal Means}\label{estimated-marginal-means}}

We performed unplanned post-hoc tests to explore differences between factor levels. We used \texttt{emmeans} \citep{lenthemmeans2023} to compute estimated marginal means at all factor levels. In \texttt{emmeans}, estimated marginal means are computed by holding all numeric covariates at their means, then averaging across a balanced grid of categorical predictors \citep{heiss2022}. To further understand differences between factor levels, we used the \texttt{contrast()} function in \texttt{emmeans} to calculate differences between marginal means. We report these estimated marginal means (EMM) and their comparisons in the following sections.

\subsubsection{Thematic Analysis}
\marked{Given the scope of this paper, our qualitative analysis focused on the following survey questions:}

\begin{itemize}
    \item What strategies have you used to reduce or prevent spending?
    \item{How have family or friends helped you prevent spending?}
    \item{What other techniques have you used to reduce the risk of impulsive spending?}
\end{itemize}

\marked{Given the exploratory nature of this initial survey, we applied a reflexive and inductive approach to thematic analysis to the open-ended responses. To uncover the themes in the insights shared, we deployed Braun and Clarke's 6-step approach \cite{braun2019reflecting} to thematic analysis.  (1) First, we familiarized ourselves with responses provided by participants. Some responses were very specific and to-the-point, consisting of only a few words (e.g., \emph{"monitoring my spending"}). Others provided lengthy responses, describing different financial activities in specific detail. (2) We then worked through the responses, highlighting different sections and applying initial labels to standout quotes. (3) From there, we extended these initial codes to the whole data set, clustering codes or creating new codes as new ideas presented in the responses to generate overall themes. (4) These themes were then reviewed and refined by multiple members of the research team. This involved combining some highly related themes and discarding others with limited representation within the data set. (5) Once these themes remained persistent with no additional findings or variation throughout the full data set, we defined these final themes to characterize the behaviors and social mechanisms that presented within this space. (6) The written results of this procedure and these themes in context of future work are detailed in the next sections.}

\begin{figure*}
\includegraphics[width=0.90\textwidth]{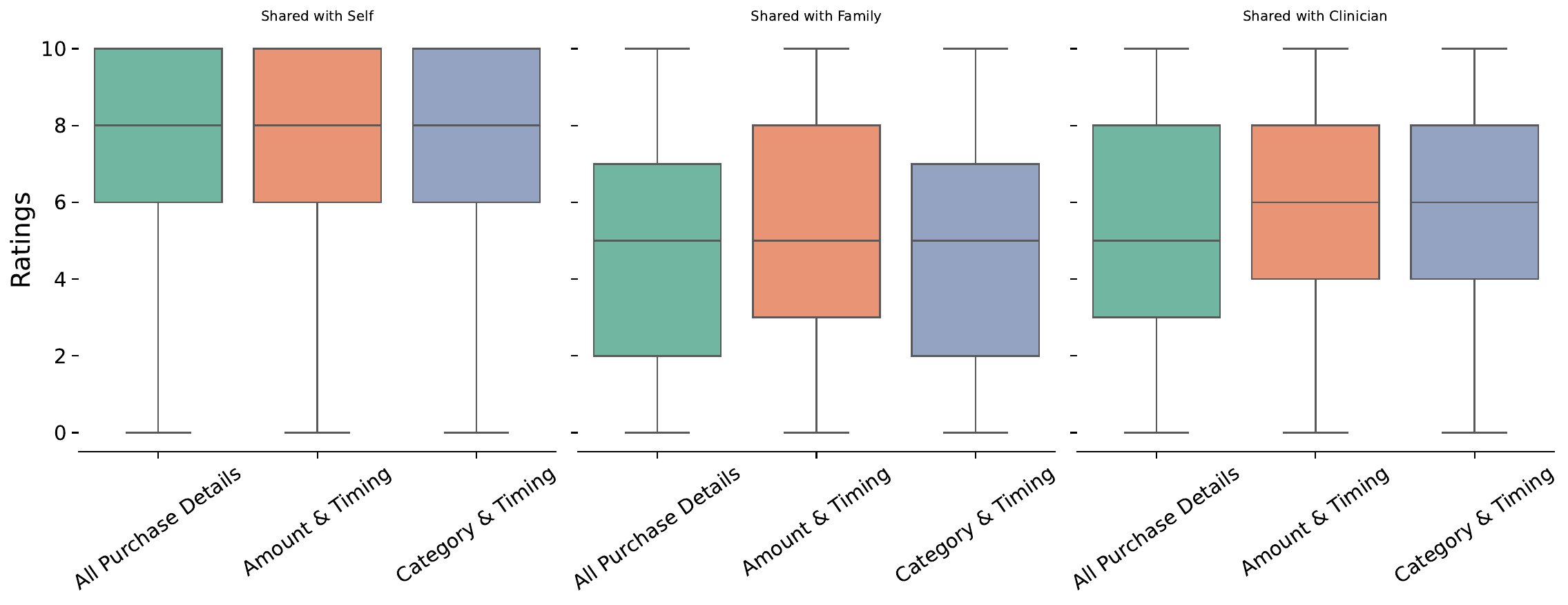}
\caption{Participants were comfortable sharing all financial data types for their own review. They were less comfortable sharing all purchase details with family (see section 4.3). We have included descriptive statistics for all vignettes in supplementary materials.}
\label{fig:overall-granularity}
\Description{A faceted categorical box plot showing comfort ratings on the Y axis and data granularity categories on the X axis. The plot is faceted by data recipient (actor). For each facet, comfort ratings are shown on a scale of 0--10 for three data granularities. Participants were most comfortable sharing for self-review across all data granularities.}
\end{figure*}

\subsubsection{Positionality Statement}
The first and second author lead the development of the survey questionnaire and analysis protocols, with the insight of other authors throughout each stage. The first author conducted the quantitative analysis of the factorial vignettes, while the second author led the qualitative thematic analysis, based on their prior experience and expertise. Both analyses were reviewed by the research team. \marked{Two authors have lived experience with bipolar disorder, which provided highly relevant perspectives and considerations in the project. This included paying special attention to how we address potentially sensitive topics, as well as gathering additional contextual information important to understanding individual experiences of BD \cite{hawke2022embedding}. Authors 1, 2, 4, and 7 all have experience designing and developing health technologies for underrepresented and stigmatized populations. The research team also included three clinicians with significant experience with bipolar disorder research and clinical practice (authors 3, 5, and 6)}. To ensure robustness and replicability, we focused on discussion and agreement among the interdisciplinary research team members.

\hypertarget{results}{%
\section{Results}\label{results}}

This section describes the outcomes of our quantitative and qualitative analysis. We first describe the participant demographics. Next, we report the norms and preferences in sharing financial data. We also identify how different contextual and demographics factors impact sharing preferences. Finally, we describe findings from our thematic analysis of open-ended survey questions involving strategies for self-management and collaborative decision-making.

\subsection{Demographics}\label{demography-description}

The survey population included individuals with different diagnostic subtypes of BD. We collected data from individuals with BD type I (38\%), bipolar disorder type II (40\%), BD not otherwise specified (14\%), and cyclothymia (6.5\%). 6.5\% of respondents did not know their diagnostic subtype. The majority of the participants (96\%) had access to bank accounts (i.e., “banked individuals”). Twenty-two percent of participants reported having declared bankruptcy due to impulsive spending during manic episodes, while 37\% reported having considered declaring bankruptcy for the same reason. In other words, 59\% of participants either have considered or actually declared bankruptcy. This is consistent with prior work on higher likelihood of experiencing bankruptcy in BD \citep{nauAssessment2023}.

The survey respondents were most likely to have completed a 4-year university degree (38\%) and to live in an urban environment (48\%). Most individuals reported being single and having never married (31\%), while 29\% reported being married with children and 13\% were married without children. The survey population included 64\% individuals who identified as female, 32\% as male, 3.1\% as non-binary, and 0.3\% preferred to not describe their gender. The majority of the survey respondents were white (88\%) and 16\% describe their ethnic background as Hispanic in origin. Respondents were from the United States (42\%), the United Kingdom (39\%), or Ireland (11\%), while 8\% did not identify their country of origin. Neither ethnicity nor country were found to be significant in our full model. We have provided detailed demographic information in supplementary materials.

\hypertarget{overall-level-of-comfort}{%
\subsection{Privacy Norms for Sharing Financial Data to Support Longitudinal BD management}\label{overall-level-of-comfort}}

Overall, participants were willing to share financial data to support longitudinal BD management. The mean rating for all vignettes was 6.1 (SD=3.12) on a 0—10 scale (0 = `extremely uncomfortable’, 10 = `very comfortable’). Figure \ref{fig:overall-granularity} shows overall distribution of privacy ratings across different recipients and data granularity. Participants were most comfortable to share data with themselves as expected. Data sharing scenarios with family members and clinicians received high average ratings as well indicating overall acceptance by our participants. We have included descriptive statistics for all vignettes in supplementary materials.

\begin{figure*}
\includegraphics[width=0.75\textwidth]{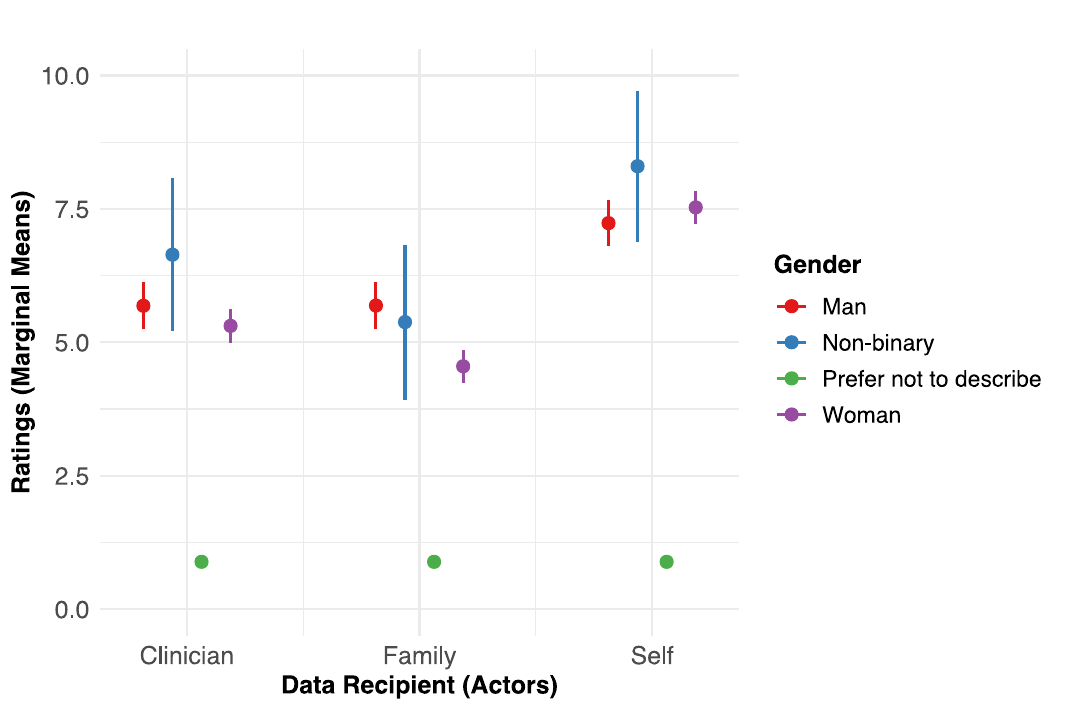}
\caption{Our data shows significant gender differences in willingness to share financial information. Women were significantly less comfortable sharing financial data with their family, although they were highly comfortable sharing data with themselves.}
\label{fig:gender}
\Description{An interaction plot showing estimated marginal means of participant comfort ratings by self-identified gender. Data Recipients (Actors) are on the X axis. Estimated marginal means of comfort ratings are on the Y axis. A legend lists "Man", "Woman", "Refer not to describe", and "Non-binary". Each legend item is displayed for each data recipient category.}
\end{figure*}

\hypertarget{factors-impacting-level-of-comfort-in-sharing-financial-data}{%
\subsubsection{Factors Impacting Financial Data Sharing Preferences}\label{factors-impacting-level-of-comfort-in-sharing-financial-data}}

Following the CI framework, we explored relative importance and impacts of contextual factors in financial data sharing preferences.


\hypertarget{actors}{%
\paragraph{Actors}\label{actors}}

Among the contextual factors, we have found actors (data recipients) to be consistently significant in determining data sharing preferences. Participants were most comfortable sharing financial data with themselves (EMM=7.44, SE=0.12). They were relatively less comfortable sharing financial data and insights with clinicians (EMM=5.44, SE=0.12) and their families (EMM=4.94, SE=0.12). Tukey pairwise comparisons showed that sharing for one's own review predicted higher levels of comfort when compared to sharing with clinicians (estimate=2.0, SE=0.06, p=.00) and sharing with family members (estimate=2.5, SE=0.06, p=.00). This implies that individuals with BD are most comfortable sharing financial data for their own review in self-management activities.

\hypertarget{contexts-of-use}{%
\paragraph{Contexts of Use}\label{contexts-of-use}}

Context of use was not a significant factor in our data set. Estimated marginal means were similar across different contexts of use scenarios — comparing mood logs to spending habits (EMM=5.96, SE=0.12), identifying distinct changes in spending (EMM=5.96, SE=.0.12), and predicting relapse (EMM=5.89, SE=0.12). No significant pairwise differences were shown in these factor levels. In other words, participants were not concerned about \emph{how} financial data might be used when it comes to supporting longitudinal BD management.

\hypertarget{financial-data-granularity}{%
\paragraph{Financial Data Granularity}\label{financial-data-granularity}}

However, granularity of financial data was an important factor in sharing preferences. Individuals were most comfortable sharing the amount and timing of transactions (EMM=6.01, SE=0.12) and slightly less comfortable sharing category and timing (EMM=5.96, SE=0.12) or all transaction details (EMM=5.84, SE=0.12). Contrast analysis for these factor levels shows a significant negative effect when sharing all transaction details (estimate=-0.97, SE=0.03, p=.01). This suggests that participants are less comfortable sharing all available transaction details. A Tukey pairwise comparison showed a significant difference between sharing all transaction details and sharing the amount and timing of transactions (estimate=-0.171, SE=0.57, p=.00). These results suggest that financial data granularity can impact sharing preferences --- our participants were more comfortable sharing the amount and timing of transactions.

\begin{figure*}
\includegraphics[width=0.75\textwidth]{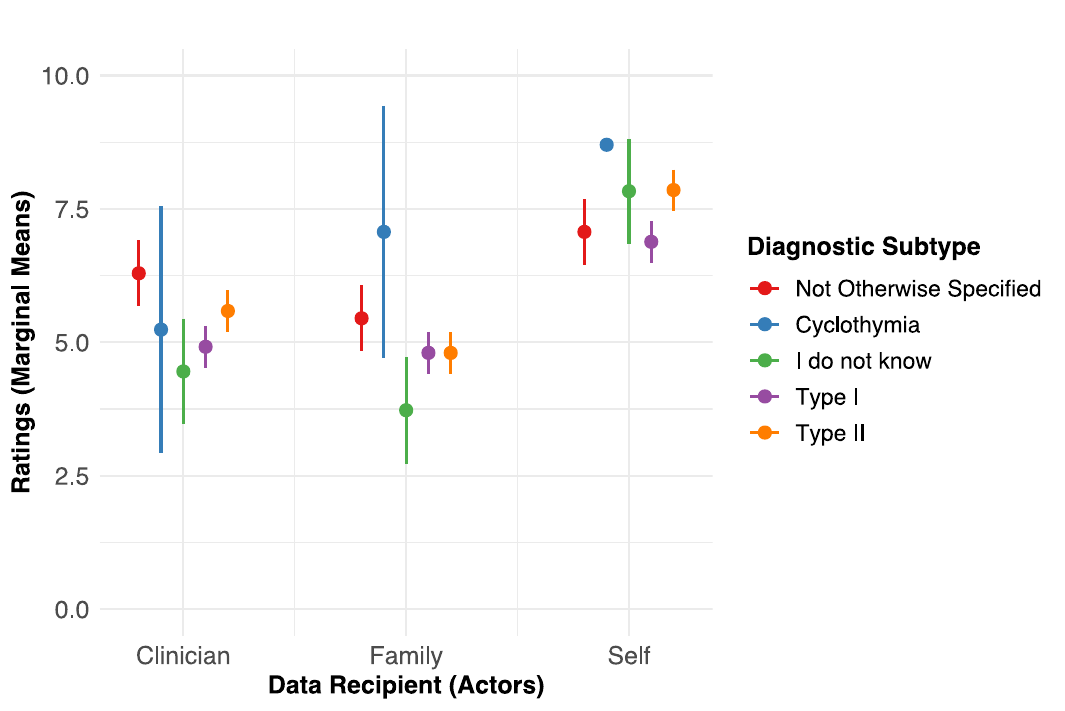}
\caption{Financial data sharing preferences varied across BD subtypes. Individuals with BD type II were significantly more comfortable sharing financial data with themselves and their clinicians compared with individuals with BD type I.}
\label{fig:diagnosis}
\Description{An interaction plot showing estimated marginal means of participant comfort ratings by diagnostic subtype. Data Recipients (Actors) are on the X axis. Estimated marginal means of comfort ratings are on the Y axis. A legend titled "Diagnostic Subtype" shows 5 items: "Not Otherwise Specified", "Cyclothymia", "I do not know", "Type I", and "Type 2". Each legend item is displayed for each data recipient category.}
\end{figure*}

\hypertarget{demographic-differences-in-comfort-level}{%
\subsubsection{Differences in Sharing Preferences Across Subgroups}\label{demographic-differences-in-comfort-level}}

We also explored how demographic differences might impact financial data sharing preferences. In our dataset, young adults (aged 18--24) were most comfortable sharing financial data with themselves (EMM=8.73, SE=0.40) and their clinicians (EMM=6.29, SE=0.40). However, they were less comfortable in sharing data with their family members (EMM=4.26, SE=0.40). Individuals in mid-life (aged 35--44) similarly reported higher levels of comfort reported in sharing data with themselves (EMM=7.76, SE=0.24) and their clinicians (EMM=5.86, SE=0.25). However, they were also more comfortable in sharing with family members (EMM=5.66, SE=0.25) compared with young adults. We conducted a contrast analysis of estimated marginal means, which further confirms the willingness to share with family members of this group to be significant (estimate=0.66, SE=0.42, p=.03). Moreover, participants were more comfortable sharing with family when they were married with children (EMM=5.76, SE=0.23) or living with a partner (EMM=5.50, SE=0.38). However, only the effect of being married with children was significant in a contrast analysis (estimate=0.96, SE=0.32, p=.03). No other marital statuses were found to be significant factors in our analysis.

We found gender differences in willingness to share financial data as well. Overall, women were comfortable sharing financial data with themselves (EMM=7.59, SE=0.15). However, they reported to be considerably less willing to share financial data with family members (EMM=4.61, SE=0.15). A Tukey pairwise comparison further revealed women were significantly less comfortable sharing with their family members than men (estimate=-1.14, SE=0.27, p=.00). These differences are visualized in Figure \ref{fig:gender}. Such gender differences in financial data sharing preferences has potentially important implications for designing supportive fintech for this population.

We also analyzed sharing preferences across BD subtypes. Individuals with BD type II were comfortable sharing financial data with themselves (EMM=7.93, SE=0.19) and with clinicians (EMM=5.66, SE=0.19). However, they were less comfortable sharing with their family members (EMM=4.87, SE=0.19). On the other hand, individuals with BD type I were in general less willing to share financial data compared with other diagnostic subtypes including with themselves (EMM=6.96, SE=0.20), with family (EMM=4.88, SE=0.20), and with clinicians (EMM=4.99, SE=0.20). There was a significant difference across BD type I and type II in willingness to share data with themselves (estimate=-0.97, SE=0.27, p=.00). This finding is consistent with prior work on the differences between BD diagnostic types. Bipolar disorder type I is associated with tendencies towards paranoia \citep{latalovaBipolar2009}, which can potentially influence their sharing preferences and privacy expectations for financial data. Figure \ref{fig:diagnosis} visualizes group differences between diagnostic subtypes across data recipients.

\hypertarget{willingness-to-accept-help-to-impede-impulsive-spending}{%
\subsubsection{Data Sharing Preferences and Willingness to Accept External Support for Financial Stability}\label{willingness-to-accept-help-to-impede-impulsive-spending}}

We investigated how willingness to use external support for financial stability might relate to data sharing preferences for individuals with BD. In our data set, participants willing to rely on technology or their creditors to prevent overspending were among the most comfortable sharing financial data. Participants who responded ``Yes, definitely'' to being willing to rely on technology to prevent overspending were highly comfortable sharing data with themselves (EMM=7.73, SE=0.21), with family (EMM=5.86, SE=0.21), and with clinicians (EMM=6.33, SE=0.21). Willingness to rely on creditors to help prevent overspending also reflected a similar pattern of high acceptance of sharing financial data with themselves (EMM=7.76, SE=0.23), with family (EMM=6.27, SE=0.23), and with clinicians (EMM=6.69, SE=0.23).

\begin{figure*}
\includegraphics[width=.75\textwidth]{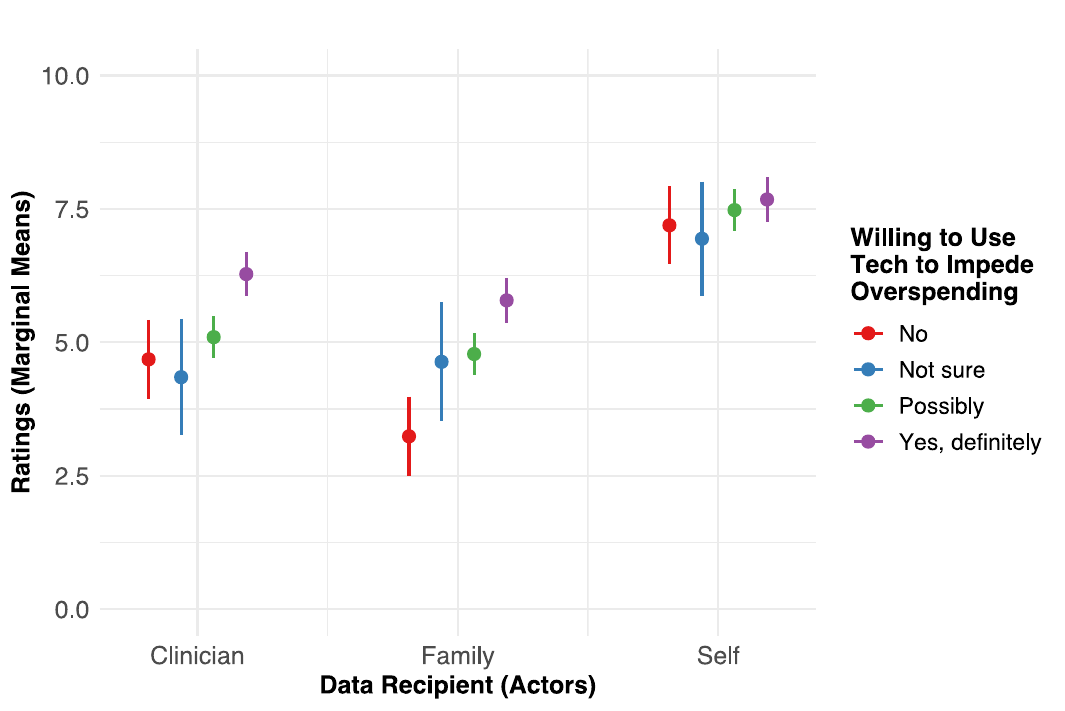}
\caption{Participants willing to adopt technology to prevent overspending were significantly more comfortable sharing financial data with clinicians and family.}
\label{fig:willingtech}
\Description{An interaction plot showing estimated marginal means of participant comfort ratings by willingness to use technology to impede overspending. Data Recipients (Actors) are on the X axis. Estimated marginal means of comfort ratings are on the Y axis. A legend titled "Willingness to Use Tech" shows 4 items: "No", "Not sure", "Possibly", "Yes, definitely". Each legend item is displayed for each data recipient category.}
\end{figure*}

These differences were significant in both cases. Those who responded ``Yes, definitely'' to technological help with overspending were more likely to share with family (estimate=1.50, SE=0.22, p=.00) and with clinicians (estimate=1.53, SE=0.22, p=.00). Those who responded ``Yes, definitely'' to help from creditors with overspending were more likely to share with family (estimate=1.18, SE=0.24, p=.00) and to share with clinicians (estimate=1.18, SE=0.24, p=.00). No significant differences were found when sharing with oneself.

Participants who were unwilling to adopt external support were significantly less comfortable in sharing financial data. Individuals unwilling to use technology to impede overspending were significantly less likely to be comfortable sharing with family (estimate=-1.37, SE=0.33, p=.00) and less comfortable sharing with clinicians (estimate=-0.42, SE=0.32, p=.26). Participants unwilling to allow creditors to impede overspending were also less comfortable sharing with family (estimate=-0.84, SE=0.25, p=.00) and with clinicians (estimate=-0.50, SE=0.25, p=.06).

The differences in these groups (i.e., between ``No'' and ``Yes, definitely'') were significant. With respect to willingness to rely on technology, a Tukey pairwise comparison showed significant differences in family sharing (estimate=-0.25, SE=0.43, p=.00) and when sharing with clinicians (estimate=-1.60, SE=0.43, p=.00). With respect to willingness to rely on creditors, a Tukey pairwise comparison showed similar patterns. Differences between these two response groups were significantly lower in family sharing (estimate=-2.32, SE=0.36, p=.00) and in clinician sharing (estimate=-2.03, SE=0.36, p=.00). Figure \ref{fig:willingtech} shows the differences in sharing preferences across these subgroups.



\subsection{Summary of Quantitative Findings}
\marked{Our findings show that the participants are comfortable in using financial data to support illness management and ensure financial stability.  We also found that participants’ data sharing preferences can depend on recipients and data granularity, while context of use was not a significant factor in our data. There was a significant gender differences in willingness to share financial data --- women were less comfortable in sharing data with their family members. BD subtypes were also associated with data sharing preferences. For example, individuals with BD type II were more comfortable in sharing data with clinicians compared with individuals with BD type I. Furthermore, individuals open to adopting fintech were significantly more comfortable in sharing financial data. These exploratory findings establish a preliminary evidence base of financial data sharing attitudes in individuals with BD.}

\subsection{Collaborative Interactions and Strategies to Sustain Financial Stability}
We conducted a thematic analysis of open-ended survey responses to better understand the current practices and challenges in collaborative financial management for individuals with BD. The majority of respondents reported involving their family or friends to help manage their spending decisions and overall finances. Specifically, half of the survey respondents reported actively involving others to help prevent impulsive spending. Participants engaged in a broad spectrum of collaborative interactions to support their financial wellbeing, ranging from disclosing financial information to relinquishing complete control to their care partners.

\subsubsection{Levels of Financial Collaboration}
Care partners \marked{sometimes provided \emph{direct financial assistance} to sustain the longitudinal financial stability of individuals with BD}. Specifically, family members might pay off debts or loan money to improve their financial situation. Such assistance might also include longitudinal monitoring and management over time by care partners. For example, P34 noted: ``\emph{my mom helped me pay off a lot of credit card debt. After that she stayed on my account so she could semi monitor and make sure I don't get into so much debt again}.” 

Participants also mentioned handing over some level of \emph{control in financial decisions} to care partners. \marked{Approximately 46\% of respondents mentioned that they had at least temporarily handed over control of their finances when recalling the management strategies they had used in the past.} Many individuals had a spouse or a parent fully managing their individual or household finances on a long-term basis. One participant commented: ``\emph{my husband takes care of the finances now. I do not have access to money”} (P460). Others called on friends or family to take over their finances during the onset of symptomatic episodes. For example, P233 noted: ``\emph{when I'm manic, I'll transfer some of my money to a family member so I don't have immediate access to it, preventing impulsive spending}.’’ Some participants also took collaborative steps to prevent overspending for specific events. As P413 commented, they ``\emph{have had friends been in charge of [their] wallet on nights out}.”  

Some participants maintain financial control, while allowing care partners to \emph{monitor their behaviors}. Additionally, a care partner may have the password to their online banking account or a copy of their credit card statement but not the permission to make or stop any transactions. This type of strategy was exemplified by P87 who had family members ``\emph{watch out for any overly generous or excessive spending}" or P442 who regularly ``\emph{sent credit card statements to [her] husband}.” \marked{Approximately 20\% of respondents reported that having a care partner monitor their finances was their primary management strategy.}

Some care partners took on the role of \emph{financial advisor}. In this case, respondents might not give direct access to banking information but share details about their finances during regularly scheduled money talks or while asking for occasional advice. During these money talks, respondents and their care partners might check in about budgeting goals and overall progress. Respondents also sought out advice or approval prior to making specific financial decisions, such as especially large purchases. For example, P375 commented: ``\emph{I find it helps if I talk to my spouse or other family/friends sometimes when I am considering a large purchase. It makes me pause and be more realistic}.” \marked{About 30\% of respondents reported turning to their care partners as a source of financial advice.}

Care partners also devised specific strategies to ensure longitudinal financial stability of individuals with BD. P305 commented that ``\emph{friends accompany me at home to make delicious food and other distractions}.” Care partners might be able to identify early warning signs and spending triggers of individuals, which can lead to proactive collaborative interactions to maintain financial wellbeing of individuals with BD. As P472 noted: ``\emph{my friends help me find alternatives when I'm really wanting an expensive new project}.” 

\subsubsection{Collaboration Strategy Can Change Over Time}
\marked{Our data set reflects the varying and intermittent needs of individuals with BD when it comes to financial collaboration. Given that symptom severity and individual needs can vary considerably across different stages in BD, respondents did not rely on one collaboration strategy but switched between them over time or used them in tandem with each other as a safety net.} This scaling up or down in care partner involvement would depend on their current need (e.g., symptom severity) and past success using these strategies. This scaling up process is exemplified by P99, ``\emph{[my] partner has access to view [my] account. [He] has [the] ability to prevent [my] access to [the] main account if needed}.” 

\marked{Several participants mentioned adopting collaborative monitoring and intervention strategies following direct financial assistance from care partners.} For example, some care partners would first provide financial assistance towards a respondent’s debt and then engage in monitoring behaviors to help them avoid having to do so again in the future. For instance, P351 commented: ``\emph{my parents paid off a credit card and then took the card for a time so I wasn't tempted to spend anything on it}". \marked{However, these collaboration strategies require ongoing negotiation over time to balance control and agency. For example, after recovering from financial instability, individuals with BD and their care partners might renegotiate the scope of collaborative monitoring and intervention to support agency and control. P40 mentioned such a transition of collaborative strategy over time: ``\emph{they helped me ring-fence money so that I can't easily spend on my own [and now] will just give honest input on big purchases}.” The need to balance agency and maintaining stability can become a source of conflict between care partners and individuals with BD as P117 mentioned: ``\emph{they keep telling me this is getting me nowhere}.”}

\subsubsection{Challenges for Financial Collaboration}
Despite regularly using these strategies, responses show that these efforts are not always successful in practice. \marked{Even when not asked directly about the challenges of collaboratively managing finances, 26\% of responses conveyed interpersonal challenges as a result of working with care partners or attempting to get help from others.}  Participants reported that they would find ways to circumvent spending restrictions and collaborative management steps. For example P73 noted: ``\emph{my husband manages our finances, but I still find ways around it}.” Furthermore, collaborative strategies between some individuals with BD and their care partners were not consistently effective as noted by P106 ``\emph{my husband double checks my bank account, tells me to stop spending. Sometimes I listen}.”

For some participants, these financial management efforts can negatively impact interpersonal relationships. Some responses noted how closely their care partners watch their spending and their overall behaviors. Some participants also mentioned the level of judgment they felt from others stemming from such surveillance. For example, P350 commented ``\emph{my spouse yells at me which does not help at all as it adds to my guilt}.” Some participants mentioned constant surveillance of their spending behaviors by care partners (P35: ``\emph{they're always looking over my shoulder}''). P451 also noted ``\emph{my husband keeps a check on my spending---watching packages arrive, watching our vacations, watching my time on computer [and] phone}.”

Some participants highlighted potential concerns regarding surveillance and losing control over their financial decision-making. For example, P187 commented: ``\emph{I didn't have a bank account so then [my] partner was meant to manage finances etc. Bad idea for all sorts of reasons [by the way]}.” Prior work has identified how surveillance can enable financial abuse \citep{belliniPaying2023}. These factors might explain our earlier findings regarding gender differences in sharing preferences --- specifically, why women in general might feel less comfortable sharing financial data with their family members.

Participants also noted the lack of care partner expertise and understanding to adequately support their needs over different stages of BD. In some cases, friends or family simply did not see the need for extra support. P450 noted her failed attempt at asking for help: ``\emph{I asked a friend to hold my debit card for me so that I could not use it without getting it from her, but she found it funny and told me to shop for myself and enjoy whatever I bought. It didn't make sense to her}.” Others had care partners who believed that managing their spending was something they should be able to do on their own. For example, P169 noted the lack of support from their partner: ``\emph{I have asked my partner to take over my finances but he won’t because he thinks it’s a form of codependency.}” Due to the lack of expertise and education, care partners may not understand the mechanisms driving these behaviors and the type of support needed to accomplish their goals. The resultant lack of support and understanding might cause further guilt and shame for individuals seeking out help.

\subsubsection{Motivation for Financial Collaboration}
The survey data also provides insights on why individuals with BD may be highly motivated to engage in collaborative financial interaction, despite the challenges it can present. To start, individuals may be in a situation where they need direct financial assistance to gain better footing before they can begin actively working toward financial stability. Specifically, individuals with BD might need such support following an episodic period, which can lead to high levels of debt or even risk of bankruptcy \citep{nauAssessment2023}. By paying off debts or loaning money, care partners provide individuals with a clean(er) slate and allow them to move toward better financial stability. Similarly, collaborative management steps including periodic monitoring can help towards longitudinal financial stability. As P124 commented: ``\emph{my husband has taken my card, blocked PayPal, and then helped me set up a budget}.” Participants also noted how collaborative interactions can lead to better informed decision-making over time. Individuals may be especially motivated to involve others when making large purchases or important financial commitments. For example, P63 commented:  ``\emph{if I'm getting too silly, mum will come shopping [with me] and be the voice [of] reason for me}.”

Lastly, individuals may see their care partners as a source of new information and an opportunity to learn new financial skills, such as budgeting strategies or spending priorities. P112 noted how their care partners ``\emph{help me ring-fence money so that I can't easily spend [it] and will give honest input on big purchases}." Friends or family members may also share their own experiences and best practices, providing a blueprint for individuals who want to establish new financial behaviors and long-term goals.

\section{Discussion}
Our survey suggests that there is a high level of acceptance of using financial data for BD management. However, some demographic groups may be more hesitant to share their financial data with others. Despite this, the majority of respondents also reported actively involving friends and family in their financial management strategies. Following these findings, we discuss the unique privacy needs of individuals with BD. We also suggest potential directions for designing supportive financial technologies for this population including self-management, financial collaboration, clinical care, and financial literacy resources to support long-term stability. We also identify challenges and needs for risk mitigation in financial technologies aiming to support individuals with BD.

\subsection{Privacy Norms and Sharing Preferences are Nuanced and Context Dependent}
BD can lead to complex and highly individualized symptoms and behavioral outcomes. The resultant privacy needs and sharing preferences for financial data are varied and context dependent. Attitudes toward financial data sharing may change over time and context, may go against support needs, and be affected by perceived risks of financial abuse and exploitation. In the following sections, we discuss factors and contexts that might impact privacy norms and sharing preferences for financial data.

\subsubsection{Changing Needs Over Time}   
Privacy norms are not static. Rather, they change over time and situational contexts. For individuals with BD, data sharing preferences may change over time in relation to their illness state, life stages, financial standing, and availability of support networks. Supportive fintech for this population must accommodate temporal and intermittent privacy norms occurring throughout the stages of illness. The role of temporal variation is significant for those living with BD \citep{moskalewiczTemporal2020}. Prior work has documented how these states might be accommodated in the design of interfaces and clinical workflows \citep{hoeferMultiplicative2021, matthewsQuantifying2017}. As such, it is important to consider how fintech can support inevitable changes in needs and preferences related to financial data sharing, disclosure, and collaboration over time for this population. 

\subsubsection{Individuals who need help the most may be among the most reluctant to seek it out}
Our findings also indicate how sharing preferences can lead to paradoxical outcomes. Specifically, individuals with BD type I were considerably less willing to share financial data to support longitudinal management. This is consistent with prior findings --- BD type I is associated with paranoia \citep{latalovaBipolar2009}. However, this lack of willingness to engage in collaborative, supportive interactions might lead to financial instability and poor decision-making over time. Indeed, individuals living with BD type I are significantly more likely to declare bankruptcy \cite{nauAssessment2023}. Reluctance to share data might also depend on the degree of financial difficulties faced by individuals. Considering the stigmatizing nature of financial hardship and debt, the resultant feelings of shame, guilt, or fear of judgment may make individuals less likely to share this information with others, even though they may stand to benefit most from additional care partner support. On the other hand, those with BD type II were more willing to share financial data to support longitudinal management. Tondo et al. \citep{tondoDifferences2022} found that individuals with BD type II were more often employed, married, had children, and reported \marked {a higher socio-economic status rating} than other diagnostic types.

These findings indicate the need for highly personalized supportive systems that can adapt to individuals’ preferences and address access barriers. For example, a series of brief, highly targeted scenarios incorporating contextual privacy factors could later inform tailored guidance. Incorporating the framework of contextual privacy into an onboarding process could result in personalized recommendations related to data sharing preferences (i.e., what data types are collected, what they can be used for) and potential approaches to collaboration. In this way, a system could ``meet users where they are at'' in terms of privacy expectations and disclosure preferences.

\subsubsection{Risk for Exploitation}  
While individuals with BD are open to engage in collaborative financial management, it is important to acknowledge the risk for exploitation when relinquishing financial control and access over to other people. Prior work has documented how technology can and has been used as a tool for financial abuse within the general population \cite{belliniPaying2023}. We note that several collaborative management practices reported by our participants---though not inherently malicious---share some similarities with Bellini’s taxonomy of how abusers might use technology to maliciously impact survivors’ finances \cite{belliniPaying2023}. While this prior work highlights the common motivations of restricting access, monitoring activities, and ``sabotaging" independence as a means of control, our findings show that similar approaches are often taken by care partners to provide support or actions actively requested by individuals with BD to help prevent impulsive spending and future debt. 

The context of BD may increase an individual's need to have other people involved in their financial decisions. It may also increase their potential risk for exploitation or financial abuse. This may help explain some of the demographic differences in our survey---in particular, that women may perceive themselves at an elevated risk of financial exploitation or abuse \cite{sharp2015review}. These risks may then be compounded by existing BD-related discrimination and mistreatment \cite{farrelly2014anticipated}, leading some individuals to prefer limiting access to their financial information to avoid potential negative outcomes. 

Exploitation risks identified in other populations can provide useful insights for system designers aiming to support financial collaboration for individuals with BD. Specifically, \marked{older adults} and individuals with dementia often need collaborative support to manage their finances. However, recent studies have identified risks as well as serious concerns of financial abuse and exploitation for this population \citep{zhang2023elder}. Future research should aim to address related concerns in BD and take steps to mitigate these risks when supporting collaborative financial interactions.

\subsection{Supporting Longitudinal Financial Management for Individuals with BD} 

\subsubsection{Self-Management of BD}
Respondents were highly comfortable having their financial data accessible for their own personal use. Therefore, intervention systems informed by financial data patterns can provide additional insights and new opportunities for BD self-management. By combining BD-related data sources with financial data, users can explore the complex relationships between mood and finances, understand the ways in which these behaviors may be interrelated, and identify unique behavioral triggers they may experience. Providing the ability to view financial and BD-related data in tandem can help users take a more holistic approach to their wellbeing and long-term stability goals.

More specifically, users can gain a better understanding of how life stressors may manifest as financial behaviors, such as impulsive spending sprees or falling behind on bill payments. Through continued use, financial data and noted behavioral patterns can act as indicators of BD status and lower some of the user burdens to manually track relevant behaviors. For instance, if a system senses problematic behavioral patterns in financial data indicative of previous mood episodes, users could be alerted to the risk of potential relapse onset. This preemptive warning would allow individuals to make arrangements, notify care partners, and follow up with their clinicians to better prepare for illness management. In other words, integrating financial data can give individuals a more complete picture of their illness trajectory and more time to adequately prepare for changes in symptoms. Having more time and information prior to onset is crucial for minimizing the impact of mood episodes and effective self-management in BD \cite{blair2023knowing}.

\marked{Future work should also identify effective design strategies to support longitudinal financial stability for this population. It is particularly important to identify design metaphors and visualization that can help individuals with BD attain their financial recovery and stability, while minimizing negative feelings or rumination. Prior research has explored holistic graphical representation and ambient displays to support personalized goal-driven behaviors \cite{han2023datahalo, murnane2023narrative}. Similar ambient approaches have shown promise in other challenging behavioral tracking use cases including substance abuse recovery \cite{jones2021consistent}. Graphical representations might also be relevant to support financial stability and recovery for this population. Furthermore, these data-driven approaches with a focus on abstract representation might support a diverse range of collaboration strategies between individuals with BD and their care partners while maintaining data control and privacy.}

\subsubsection{Improving Financial Decision-Making and Literacy}
Future intervention design should support developing new skills and habits to sustain financial stability. It can be particularly useful to address financial anxiety, which can often lead to counter-productive outcomes including avoidance behaviors. We believe it will be highly beneficial to design personalized interventions focusing on financial wellbeing and stability for individuals with BD. Such interventions can leverage existing clinical practices, including cognitive financial behavioral therapy (CFBT) \cite{nabeshima2015cognitive} and dialectical behavior therapy (DBT) \cite{richardsonFinancial2021, van2013randomized}, and mindfulness activities. These interventions can be particularly effective in addressing guilt and shame associated with financial loss and instability following illness episodes in BD. 

Future fintech systems can also provide users with opportunities to improve overall financial literacy. Data-driven and personalized financial literacy tools could help users set more appropriate and manageable goals, establish new money management routines, and address financial anxieties. In our survey, some individuals noted that they seek financial knowledge from their support networks. However, there is an opportunity to create personalized support systems that can generate resources tailored to address individual financial habits and needs. \marked{Additionally, individuals and care partners engaging in financial literacy training together may help foster more situational understanding and improve collaborative interactions.}

\subsubsection{Enabling Collaborative Financial Interactions and Support}

\marked{Our findings indicate that the need for and acceptance of collaborative support can vary across individuals and over different stages of BD. Supportive fintech for this population, thus, must be flexible to address different needs of individuals as well as accounting for dynamic changes in collaboration strategies over time. Specifically, future design should aim to meet users where they are regarding their privacy and disclosure norms. Designers can use vignette-based approaches similar to this study during system onboarding to determine users' initial attitudes. The resultant data could provide tailored recommendations and strategies based on their unique preferences --- allowing them to determine what financial information is shared, with whom, and under what circumstances. As these privacy preferences may change over time, a system should also provide options to include care partners or clinicians in later stages of use. Similarly, users should be able to reassign care partner roles and permissions as their needs and comfort-levels change.}


\marked{Collaborative financial management can be challenging.  It can lead to tensions and fraught relationships, even risks of domestic violence \cite{labrum2016factors}. Future fintech design should specifically aim to ease these conflicts. Furthermore, it is crucial that future work explores the delicate balance between supportive control and financial exploitation to better protect users with BD. Future design should be mindful of these risks, minimally restrictive in relation to symptomatic need, and include adequate safeguards against potential misuse of these financial technologies. Prior work on financial exploitation in other marginalized communities \cite{zhang2023elder} can be particularly useful in designing systems that can mitigate such risks for individuals with BD. Moreover, clinicians could also employ screening measures to assess potential risk for financial exploitation within their support networks \cite{greene2022elder}. Overall, fintech design for long-term stability should prioritize positive communication skills, focusing on attainable goal setting, work to reduce feelings of financial shame and guilt, and help care partners gain a stronger understanding of the needs and experiences of those with BD. The combination of mental health and financial data comes with increased privacy risks. Moving forward, it is important to focus on developing features that support healthy ``financial collaboration” rather than creating system structures that inadvertently enable malicious ``financial control” of already marginalized users.}

\subsubsection{Integration with Existing Clinical Practices}
\marked{Overall, participants were relatively comfortable sharing data and financial patterns with clinicians. This provides an opportunity to use financial behavioral data within clinical settings to develop more comprehensive BD care plans \cite{blair2020opportunities}}. To start, patients and clinicians can target ``financial stability" as a specific treatment goal and measure progress through individualized financial behaviors. This is especially important for long-term stability considering the cyclical effects between mood and financial difficulties \cite{richardsonrelationship2013}. By addressing financial challenges and money-related stress, patients may more easily achieve other mood-related goals in their overall care plan.

Moreover, clinicians can also incorporate financial data alongside existing, validated clinical measures for effective decision-making. It will be critical to ensure transparency for effective integration with existing clinical practices. Specifically, the type and granularity of data shared with clinicians should be determined based on individuals’ comfort level, their unique needs, and the specific behavior change goals. For example, if impulsive spending bursts are highly characteristic of an individual’s mood status, they may share transaction timing data with their clinicians rather than all of their purchasing details. This could help individuals and clinicians set and monitor goals targeting impulsive spending while preserving privacy. Conversely, individuals could share all purchase details within their financial data if their main financial goal is to gain a better understanding of the relationship between their mood and spending, as well as uncover potential behavioral triggers that can affect long-term BD management more broadly. Overall, we believe current clinical practices and decision-making will significantly benefit by integrating financial behavior data.

\subsection{Limitations}
Our study is not without limitations. First, given the dearth of literature surrounding this topic, our quantitative analysis was exploratory in nature. Given these initial insights, future research can take a hypothesis-driven approach to investigate the specific factors that impact willingness to adopt supportive financial technologies and share this data with care partners. \marked{The survey vignettes were presented to participants in a partially randomized order to reduce response fatigue. While we didn't find any order effects, future work should aim to replicate these findings in a fully randomized survey design.} 

Despite the large sample of individuals with BD used in this study, respondents leaned heavily white and female regarding ethnicity and gender. While the insights gathered through this work may not speak universally for all individuals with BD, population-level diagnosis rates show that individuals diagnosed with BD are more likely to be demographically white and female \cite{shippee2011differences}. We distributed this survey internationally. However our sample was limited to higher income countries in North America and Europe. Therefore, our findings may not reflect the attitudes, needs, or experiences of those in lower and middle income countries. Additionally, this sample may not take into account other cultural norms and familial customs that may play a role in how individuals may involve care partners in their financial decisions or their attitudes toward sharing financial information with others. Lastly, 96\% of our respondents reported having access to a bank account. Therefore, our findings may not represent ``unbanked" individuals---those who fall through the cracks of the formal economy or choose not to use traditional banking institutions. However, prior research has demonstrated how digital finance systems can still help support traditionally unbanked individuals \cite{ibtasam2017exploration}. Further research is needed to understand the needs and attitudes of individuals who are unbanked and living with BD. 

\section{Conclusion}
This paper focuses on understanding financial data sharing preferences to support longitudinal management of bipolar disorder. Our findings show that individuals with bipolar disorder were willing to share financial data for personalized support and insights. However, we also identified significant demographic differences in sharing preferences across gender, age, and marital status. Specifically, women, younger adults, and those with more severe subtypes of BD may be reluctant to share financial data with others. Based on these insights, we discussed the unique privacy needs that should inform the future development of fintech systems to support higher-risk user groups. Finally, we have provided design suggestions for how financial behavioral data can be integrated into BD self-management and clinical practices, inform financial decisions and literacy resources, and improve collaboration with care partners. We consider this work to be a crucial step  toward developing personalized, privacy-aware systems to support financial stability and long-term BD care.

\begin{acks}
Research reported in this publication was supported in part by the National Institutes of Health's National Institute of Mental Health under award number R21MH131924 and by the National Science Foundation Graduate Research Fellowship Program under Grant No. DGE1255832. The content is solely the responsibility of the authors and does not necessarily represent the official views of the National Institutes of Health or the National Science Foundation.
\end{acks}


\bibliographystyle{ACM-Reference-Format}
\bibliography{chi.bib}

\end{document}

%% file: tables/factors.tex
\begin{table}[h]
\caption{The vignette design following the Contextual Integrity (CI) framework \citep{nissenbaumPrivacy2004}. We selected 3 factors each containing 3 levels resulting in a total of 27 vignettes.}
\label{tbl-factors}
\begin{tabular}{@{}ll@{}}
\textbf{Factor}                                              & \textbf{Level}                                                   \\ \hline
\begin{tabular}[c]{@{}l@{}}Actor \\ (Recipient)\end{tabular} & 
\begin{tabular}[c]{@{}l@{}}You\\ Family\\ Clinician\end{tabular} \\ 
\hline
\begin{tabular}[c]{@{}l@{}}Context of use\end{tabular} &
\begin{tabular}[c]{@{}l@{}}To predict relapse\\ To compare mood logs with spending behavior\\ To identify distinct changes in spending\end{tabular} \\ 
\hline
\begin{tabular}[c]{@{}l@{}}Granularity \\ (Data type)\end{tabular} &
\begin{tabular}[c]{@{}l@{}}Purchase timing and amount\\ Purchase timing and category\\ All purchase details\end{tabular} \\ \bottomrule
\end{tabular}
\end{table}